# Thermodynamics of proton insertion across the perovskite-brownmillerite transition in $La_{0.5}Sr_{0.5}CoO_{3-\delta}$

Armand J. Lannerd[a], Nathan J. Szymanski[a], and Christopher J. Bartel[a,*]

[a]University of Minnesota, Department of Chemical Engineering and Materials Science, Minneapolis, MN 55455

*correspondence to cbartel@umn.edu

**Abstract**

$La_{1-x}Sr_xCoO_{3-\delta}$ is a promising off-stoichiometric metal oxide that undergoes a topotactic perovskite ($\delta = 0$) to brownmillerite ($\delta = 0.5$) transition under electrochemical and thermochemical stimuli, with concomitant variations in its electrical, magnetic, thermal, and optical properties. Recent studies on thin-film cycling in electrochemical devices show incomplete reversibility of this transition, with significant acid-etching serving as a degradation mechanism. While earlier investigations examined the protonation of brownmillerite $SrCoO_{2.5}$, the thermodynamics of protonation across the perovskite-to-brownmillerite transition remain poorly understood. In this work, we combine density functional theory calculations with predictions from universal machine-learning interatomic potentials to elucidate the energetics and implications of protonation across the transition for $La_{0.5}Sr_{0.5}CoO_{3-\delta}$. These calculations reveal negative hydrogen insertion energies and strong competition with oxygen vacancy formation across the transition for a wide range of conditions. The extent of protonation is primarily limited by the availability of Co $3d$ states to accommodate reduction by inserted hydrogen. Although hydrogen insertion is often thermodynamically favorable within a defect picture, a convex hull analysis of the resulting $H_yLa_{0.5}Sr_{0.5}CoO_{3-\delta}$ phases reveals them to be unstable against decomposition into hydroxides among other products. This instability increases with hydrogen content and provides a thermodynamic basis for the acid-etching observed during electrochemical cycling. In the absence of decomposition, our calculations predict protonation to cause significant structural expansion and widening of the band gap. This work advances the fundamental understanding of protonation in $La_{0.5}Sr_{0.5}CoO_{3-\delta}$ and contextualizes experimental observations of related materials in the presence of moisture or $H_2$.

# I. Introduction

Some metal oxides can accommodate substantial off-stoichiometry, hosting non-dilute concentrations of vacancies and other point defects. [1,2] The tuning and control of these defect populations underpins a number of emerging technologies in energy and electronics. Energy-related applications include the production (e.g., solar thermochemical water-splitting) and consumption (e.g., solid oxide fuel cells) of renewable fuels. [3,4] For electronics, the last decade has seen a surge of interest in electrochemical devices, particularly electric double-layer transistors (EDLTs). [5–9] The construction of EDLTs is similar to that of a metal-oxide-semiconductor field-effect transistor, except the metal oxide dielectric has been replaced by an electrolyte (e.g., an ionic liquid or gel) and the semiconductor, in many cases, by an off-stoichiometric metal oxide. During electrolyte gating, the electrolyte-oxide interface acts as a nanometer-scale parallel-plate capacitor with a large specific capacity, leading to significant modulation of charge carrier densities and off-stoichiometry, thereby enabling control over a wide range of physical properties in the oxide. [5,7,10]

The perovskite $La_{1-x}Sr_xCoO_{3-\delta}$ (LSCO) system has been investigated extensively across both domains, showing particular promise in EDLTs. [5,11–22] LSCO exhibits a strong electrochemical response to an applied gating voltage ($\eta$) due to its relatively low oxygen vacancy ($v_O$) formation energies and high $v_O$ diffusivities. [23] Under a positive bias, LSCO is reduced via $v_O$ formation, with $O^{2-}$ ions passing out of the oxide and into the electrolyte across the electrolyte-oxide interface. Above a threshold voltage, $v_O$ ordering occurs, eventually realizing a topotactic phase transition from perovskite (P) $La_{1-x}Sr_xCoO_{3-\delta}$ to brownmillerite (BM) $La_{1-x}Sr_xCoO_{2.5}$. In the P phase with $\delta = 0$, Co is octahedrally coordinated with O. As $\delta$ increases (but remains below 0.5), disordered $v_O$ lower the average Co-O coordination. At $\delta = 0.5$, the BM phase has alternating octahedral $CoO_6$ and tetrahedral $CoO_4$ layers along the [001] direction resulting in 1D oxygen vacancy channels. Oxidation from the BM phase to the P phase occurs under a negative bias. There is strong evidence that residual $H_2O$ present in the electrolyte serves as an oxygen reservoir, with $H_2O$ electrolysis playing a key role in the oxidation process. [5,12,23,24]

The topotactic P-to-BM transition in LSCO is accompanied by a metal-to-insulator transition (MIT, for $x > 0.18$) [15,19,21], a change from ferromagnetic (FM) to weakly FM or G-type antiferromagnetic (G-AFM) ordering [14,15,20,24], wide modulation of 300 K thermal conductivity [16,25], and large changes in complex refractive index within the visible and infrared range [17,24,26]. Reversible modulation between these states opens pathways to applications in resistive switching for neuromorphic computing [27], magnetoionics for next-generation memory [28], thermal switches for temperature regulation [29], visible-range electrochromics [30], and more [5].

Reversibility and extended cycling of the P-to-BM transformation remains an outstanding challenge for most practical applications of LSCO. A recent study carried out a gate-voltage hysteresis loop analysis, where ultrathin (10-unit-cell-thick) epitaxial $La_{0.5}Sr_{0.5}CoO_{3-\delta}$ was fully cycled once in an EDLT device under vacuum at 300 K. [12] While the P-to-BM-to-P transformation proceeded with near complete reversibility, a slight discontinuity between the measured source-drain current of the initial and final P phases indicated small irreversible changes between these states. Additional observations suggested that acid-etching had likely occurred, as the overall film thickness decreased and surface roughness increased. [12] It has been speculated that protons generated via electrolysis of the residual $H_2O$ at high positive voltages are the acid source. These acid-etching effects have been noted previously for LSCO and CdO operating under similar conditions. [20,31] In another recent work, the cycling endurance of thin film LSCO was investigated via operando FTIR to track the optical MIT transition behavior with cycling. [26] Under optimized conditions, ~40 cycles of full optical MIT transitions were achieved, with additional limited switching behavior up to 100 cycles. Critically, acid-etching of the LSCO film, especially near the metal contacts, was identified as the primary degradation mechanism. Both the choice of noble metal for the contacts and the relative humidity heavily affected the degree of acid-etching, with higher relative humidity exacerbating acid-etching. Concurrently, the oxidation threshold voltage for the BM-to-P transition became more positive with increasing relative humidity, indicating less work required to oxidize in the presence of moisture. Taken together, these findings point to higher $H_2O$ concentrations in the electrolyte having a significant influence on the cycling performance of LSCO.

Observations of acid-etching and the direct effect of relative humidity suggest that there are underexplored interactions arising when LSCO is in the presence of water or electrolysis products (such as hydrogen). One such interaction may be the formation of hydrogen interstitials ($H_i$) giving rise to protonated LSCO phases. A previous electrolyte gating experiment established that after gating $SrCoO_{3-\delta}$ to $SrCoO_{2.5}$, it was possible to further reduce the material by gating from $SrCoO_{2.5}$ to $HSrCoO_{2.5}$. [24] $SrCoO_{3-\delta}$ and $HSrCoO_{2.5}$ were found to decompose into $SrCoO_{2.5}$ above ~195 °C and ~175 °C, respectively. The consequences of protonation were structural expansion, widening of the band gap, and activation of a weak FM state (instead of the G-AFM state associated with $SrCoO_{2.5}$). Notably, structural expansion and widening of the band gap are the same effects observed for oxygen vacancy formation, making it difficult to disentangle hydrogen insertion from oxygen removal. In addition to several indirect indicators, the presence of inserted hydrogen was directly confirmed via secondary-ion mass spectrometry. Subsequent experimental works demonstrated that $H_{1.5}SrCoO_{2.5}$ and $H_2SrCoO_{2.5}$ phases could be accessed under extreme gating conditions through the formation of charge-neutral H-H dimers. [32,33] Additionally, $HSrCoO_{2.5}$ was found to demonstrate high proton conductivity, showing promise as an electrolyte for low-temperature direct hydrogen solid oxide fuel cell applications. [34] There are also reports of a similar $H_ySrFeO_{2.5}$ phase. [35]

Several computational investigations of $HSrCoO_{2.5}$ followed to understand the effect of hydrogen interstitials on magnetic properties [36–38] and the origin of high proton diffusivity [39,40]. Across all studies, the hydrogen interstitials were found to form hydroxyl (−OH) bonds, where the hydrogen is positively charged. Interestingly, multiple studies reported negative hydrogen interstitial formation energies for both dilute ($y \approx 0$) and non-dilute ($y \gg 0$) hydrogen stoichiometries. [38–40] This suggests that hydrogen insertion is thermodynamically favorable in the BM $SrCoO_{2.5}$ phase. The plausibility of hydrogen insertion on mixed La/Sr cobaltites (i.e., $x < 1$ in $La_{1-x}Sr_xCoO_{3-\delta}$) is less studied. [41,42] There is one report of $H_{0.67}La_{0.67}Sr_{0.33}CoO_{2.67}$ formation upon electrolyte gating of $La_{0.67}Sr_{0.33}CoO_{3-\delta}$ at slightly elevated temperatures. [41] Another report suggests protonated P and BM phases of $La_{0.7}Sr_{0.3}CoO_{3-\delta}$ can form upon thermochemical annealing in a $H_2$ atmosphere. [42] Both studies observe structural expansion and increasing insulating behavior as a consequence of protonation. Accompanying calculations also suggest once more that the hydrogen interstitials exist as hydroxyl bonds and that hydrogen insertion is facile owing to negative hydrogen interstitial formation energies. [41,42]

In this work, we seek to expand our understanding of proton insertion in $La_{0.5}Sr_{0.5}CoO_{3-\delta}$ through a comprehensive first-principles thermodynamic analysis across the P-to-BM transition. We examine the competition between oxygen vacancy and hydrogen interstitial formation and explore the limiting factors and consequences of protonation. Taken together, our findings lend insights into the acid-etching degradation mechanism previously discussed for EDLTs and may also have implications for the use of LSCO in other technologies such as solid oxide fuel cells.

## II. Methods

### A. First-principles calculations

Density functional theory (DFT) calculations were performed to determine the energetics of oxygen vacancy and hydrogen interstitial formation across the P-to-BM transition, to assess the stability of the resulting structures, and to evaluate the consequences of defect formation on the structural and electronic properties of LSCO.

All DFT calculations were carried out via the Vienna *Ab Initio* Simulation Package (VASP, version 6.4.1) [43,44] using the projector augmented wave (PAW) [45,46] method. The exchange-correlation functional was approximated by the Perdew-Burke-Ernzerhof (PBE) [47] generalized gradient approximation (GGA) with an effective Hubbard U correction (GGA+U) to account for spurious self-interaction effects. [48] Consistent with earlier studies on LSCO, we employed an effective Hubbard value of 3.0 eV for Co. [41,42,49,50] All calculations had a plane-wave energy cutoff of 520 eV, a Γ-centered k-point grid with a minimum spacing between k-points of 0.22 Å$^{-1}$ (KSPACING in VASP), and were spin-polarized. For each unique structure, Co was initialized with a MAGMOM value of +/− 1.9 $\mu_B$, where FM, G-AFM, and A-AFM orderings were considered. [19,51] The lowest post-relaxation energy of these three magnetic initializations was used for thermodynamic analysis. Convergence criteria for all calculations was set to 0.03 eV Å$^{-1}$ for ionic relaxation and $10^{-6}$ eV for electronic optimization.

It has previously been found for $SrCoO_{3-\delta}$ that the FM, G-AFM, and A-AFM magnetic orderings lie very close to each other energetically. [52] Given this, our electronic structure analyses primarily focused on assessing qualitative trends across all initializations. The LOBSTER package [53] was used for density of states analysis, and charge analysis was done via the Bader method [54,55]. Structures were visualized using VESTA. [56] Pymatgen was used to prepare and analyze DFT and LOBSTER calculations. [57,58]

**B. Universal machine learning interatomic potential calculations**

Universal machine learning interatomic potentials (uMLIPs) have emerged in recent years as powerful tools that achieve near-DFT accuracy while being several orders of magnitude faster than comparable DFT calculations. [59] Owing to the vast configurational space presented by non-dilute defect structures in the $H_yLa_{1-x}Sr_xCoO_{3-\delta}$ (HLSCO) system, we utilized uMLIPs in multiple instances to optimize structures and evaluate the energies of a wide range of defect configurations.

Specifically, we employed the Universal Models for Atoms (UMA) framework, using the *uma-s-1* model with the OMat task expert. [60] The *uma-s-1* model demonstrates state-of-the-art performance in structure optimization and energy prediction, as well as faster inference times compared to similarly accurate models like *eSEN-30M*. [60,61] The model was trained on both the Alexandria [62] and OMat24 [63] datasets, which include many non-equilibrium structures. Such training is critical for learning the potential energy surface far from equilibrium, an important feature for defect calculations. [64] Ionic relaxations with UMA were implemented using the Atomic Simulation Environment (ASE) [65] python toolkit, using the FIRE optimization algorithm [66] and FrechetCellFilter. The force convergence criteria was set to 0.03 eV Å$^{-1}$ and the energy of each optimized structure taken from the final inference step of relaxation.

**C. Defect thermodynamics**

The topotactic P-to-BM transition of LSCO initially involves the formation of disordered oxygen vacancies, before an eventual disorder-order transition with the emergence of the BM phase. Taking a particular host composition and structure (e.g., perovskite $La_{0.5}Sr_{0.5}CoO_3$), the energetic cost of forming a (neutral) defect (e.g., an oxygen vacancy) can be calculated from first principles by:

$$\Delta E_{\text{defect}} = E_{\text{defect}} - E_{\text{host}} + \sum_i \Delta n_i \mu_i \qquad (1)$$

In Equation 1, $\Delta E_{\text{defect}}$ is the defect formation energy, while $E_{\text{defect}}$ and $E_{\text{host}}$ are the total energies of the structure containing the defect and the pristine structure, respectively, as determined from DFT (or uMLIP) calculations. The final summation term accounts for the energy of all atoms added to or removed from the environment when introducing the defect. In this term, $\Delta n_i$ is the number of atoms added to the environment and $\mu_i$ the corresponding chemical potential, which is defined as:

$$\mu_i = \mu_i^{\text{DFT}} + \mu_i^{\text{corr}} + \Delta\mu_i(T, p_i, \eta) \qquad (2)$$

In Equation 2, $\mu_i^{\text{DFT}}$ is the 0 K reference energy calculated from DFT and $\mu_i^{\text{corr}}$ any relevant correction to that energy. The $\Delta\mu_i$ term captures temperature ($T$), concentration ($p_i$), and electrochemical potential ($\eta$) effects. In this work, the two defects of interest are v$_O$ and H$_i$. For v$_O$:

$$\mu_O^{\text{corr}} = +0.687 \text{ eV} \tag{3}$$

$$\Delta\mu_O = \mu_O(T, p^0) + \frac{1}{2} k_B T \ln\left(\frac{p_{O_2}}{p_{O_2}^0}\right) - 2\eta e \tag{4}$$

The $\mu_O^{\text{corr}}$ value of +0.687 eV is a well-known correction for oxygen overbinding in GGA calculations that was used when analyzing the DFT calculations performed in this work. [67] The training data for UMA employed slightly different DFT settings, leading to a $\mu_O^{\text{corr}}$ of +0.657 eV that was utilized when analyzing UMA calculations. [63,68] In Equation 4, $\mu_O(T, p^0)$ is the temperature-dependent Gibbs free energy from experiment at standard pressure ($p_{O_2}^0 = 1$ bar) [69], while the second term accounts for the shift in oxygen chemical potential from changes in concentration, assuming an ideal-gas (solution) behavior. In this term, $k_B$ is the Boltzmann constant and $p_{O_2}$ the oxygen gas partial pressure. The third term is the shift in oxygen chemical potential at equilibrium under an applied electrochemical potential ($\eta$), following from the Nernst equation. [70] Similarly, for H$_i$:

$$\mu_H^{\text{corr}} = -0.3508 \text{ eV} \tag{5}$$

$$\Delta\mu_H = \mu_H(T, p^0) + \frac{1}{2} k_B T \ln\left(\frac{p_{H_2}}{p_{H_2}^0}\right) + \eta e \tag{6}$$

The $\mu_H^{\text{corr}}$ value of -0.3508 eV is a correction tying the DFT calculated energy for hydrogen to the 300 K experimental formation energy for water. [71] When analyzing UMA calculations, the slightly different correction of -0.3426 eV was applied. [63,68] The correction was included to reflect the important role of residual water in the ion gel of LSCO in gating experiments. The effect of temperature, hydrogen concentration ($p_{H_2}$, with $p_{H_2}^0 = 1$ bar), and applied electrochemical potential on hydrogen chemical potential are captured in Equation 6. Importantly, although the sign of the last term is flipped between Equations 4 and 6, a positive $\eta$ will lower the relevant defect formation energy in both cases since v$_O$ involves the removal of an atom from the host to the environment ($\Delta n_O > 0$), and H$_i$ involves the addition of an atom to the host from the environment ($\Delta n_H < 0$). This is rooted in the reducing nature of both neutral v$_O$ and H$_i$.

In general, a negative $\Delta E_{\text{defect}}$ indicates the host structure is unstable with respect to the formation of that defect under the specified conditions. For positive values, entropy effects will still yield non-zero defect concentrations at finite temperature. Smaller positive $\Delta E_{\text{defect}}$ lead to higher equilibrium defect concentrations.

It is also useful to consider the interaction energy between defects at non-dilute concentrations. For a structure containing $v_O$ and $H_i$, the interaction energy is defined as:

$$\Delta E_{\text{int}} = \Delta E_{\text{defect}}(n_{v_O}, n_{H_i}) - n_{v_O}\Delta E_{\text{defect}}^{v_O}(n_{v_O} = 1) - n_{H_i}\Delta E_{\text{defect}}^{H_i}(n_{H_i} = 1) \quad (7)$$

$\Delta E_{\text{defect}}(n_{v_O}, n_{H_i})$ is the defect formation energy associated with the structure containing multiple defects ($n_{v_O}$ and $n_{H_i}$), while $\Delta E_{\text{defect}}^{v_O}(n_{v_O} = 1)$ and $\Delta E_{\text{defect}}^{H_i}(n_{H_i} = 1)$ are the dilute oxygen vacancy and hydrogen interstitial formation energies. $\Delta E_{\text{int}}$ quantifies the scale of defect-defect interactions at the defect concentrations considered. A negative value indicates that the presence of a defect makes further defect formation more favorable, whereas a positive value signifies subsequent defect formation becomes less favorable.

## D. Stability analysis

While defect thermodynamics describes the energetics of forming a particular defect, it does not explicitly capture the thermodynamic stability of a phase against decomposition into a mixture of competing phases. Stability against decomposition is often called hull stability, as it is assessed via the convex hull formalism. In this formalism, the formation energy, $\Delta E_f$, of all relevant phases is calculated as:

$$\Delta E_f = E - \sum_i n_i \mu_i \quad (8)$$

In Equation 8, $E$ is the internal energy of the compound of interest (from DFT or uMLIP calculations) with an appropriate compatibility scheme [72] applied for mixing GGA and GGA+U calculations. The 0 K DFT chemical potentials ($\mu_i$) of its constituent atoms ($n_i$) are subtracted to obtain the formation energy ($\Delta E_f$). A negative $\Delta E_f$ indicates that the compound is stable against decomposition into its constituent elements. To determine hull stability, the formation energies of all compounds considered as decomposition products are computed, with the lower convex envelope in formation energy-composition space defining the stable compounds. [73] The hull energy, $\Delta E_{\text{hull}}$, quantifies the energy above the hull of any unstable compounds. Although many metastable phases are experimentally accessible, the vast majority of experimentally synthesized materials fall within 100 meV/atom of the hull. [74]

When evaluating the hull stability of HLSCO, we investigated two scenarios. In the first, we only considered compounds in the $H_yLa_{0.5}Sr_{0.5}CoO_{3-\delta}$ family of structures to assess stability within P- and BM-derived structures (the "topotactic" hull). For the second case, we investigated the hull stability of HLSCO against all stable materials in the H-La-Sr-Co-O chemical space reported in the Materials Project database along with all $H_yLa_{0.5}Sr_{0.5}CoO_{3-\delta}$ structures calculated in this work (the "complete" hull). [75,76] To factor in the entropic stabilization of $A$-site disorder and defects in the HLSCO structures at finite temperatures, the 300 K configurational entropy of $A$-site mixing and their defects was included, assuming ideal mixing entropy.

## E. Structure search

We started from the lowest energy structures for δ = 0, 0.125, 0.25, 0.375, and 0.5 reported in Ref [49] for $La_{0.625}Sr_{0.375}CoO_{3-\delta}$ and adapted them from $x = 0.375$ to $x = 0.5$. Many thin film experiments only explicitly report the P (δ = 0) and BM (δ = 0.5) phases [12,15], while a δ = 0.25 phase has been suggested to exist in the bulk for various $x$ in $La_{1-x}Sr_xCoO_{3-\delta}$ [51,52,77,78]. Given this, for simplicity we only considered δ = 0, 0.25, and 0.5. A special quasirandom structures algorithm [79] was used to mimic *A*-site disorder in a finite cell with occupation of 50% La and 50% Sr in alignment with a lack of experimentally observed *A*-site ordering. [51]

For defect calculations, the *doped* package [80] was used to generate an ideal supercell transformation for the P phase by optimizing for the dilute defect case (maximizing the distance between periodic images) for an 80 atom cell. This same transformation was applied to the δ = 0.25 (76 atom) and BM (72 atom) structures. The resulting structures were fully relaxed via DFT and constitute three host structures for defect calculations. They are shown in **Figure 1**.

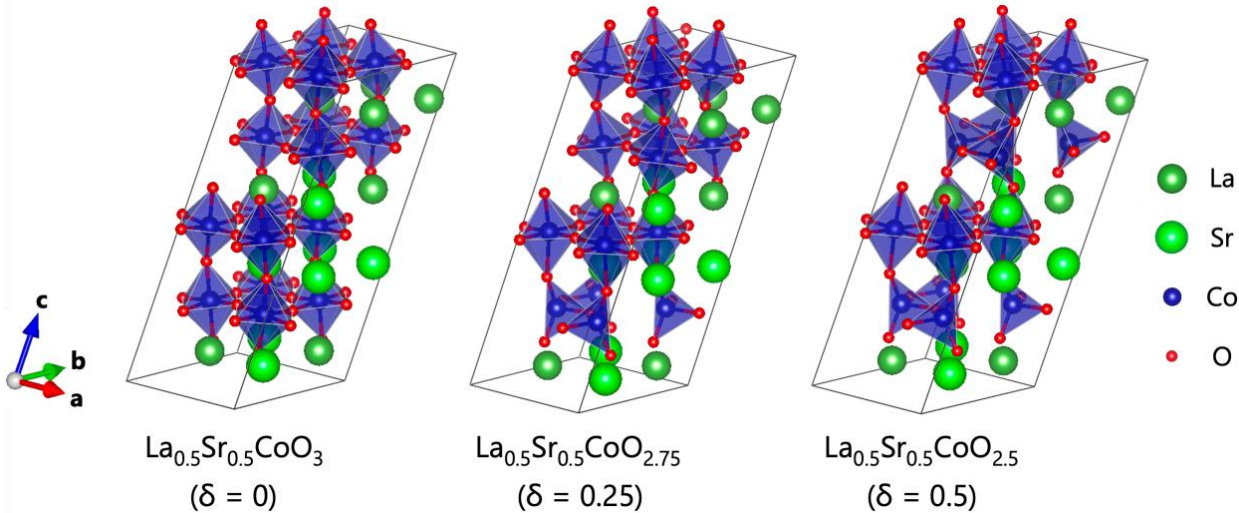

**Figure 1.** Supercell representations across the topotactic transition of $La_{0.5}Sr_{0.5}CoO_{3-\delta}$ from perovskite (δ = 0) to brownmillerite (δ = 0.5). These three structures were utilized as the host structures for defect calculations across the transition. Dark green spheres represent La, light green spheres represent Sr, dark blue spheres represent Co, and red spheres represent oxygen.

With the supercell host structures set, dilute defect configurations with a single $v_O$ or $H_i$ were generated for each host. For the dilute $v_O$ structures, all single $v_O$ were enumerated and evaluated for uniqueness using StructureMatcher from pymatgen with default settings ignoring the *A*-site disorder. [57] These structures were relaxed using DFT. The dilute $H_i$ structures were constructed in two ways. In the first, all structures were enumerated where a single $H_i$ was placed 0.9 Å away from the unique oxygen identified when generating the dilute $v_O$ structures, with the placement maximizing distance from other nearby bonds. These structures were relaxed using DFT. In the second method, *doped* was used to identify all unique interstitial sites, each yielding a unique dilute $H_i$ structure. [80,81] These structures were first relaxed with UMA, with the lowest-energy UMA structure at each composition subsequently relaxed using DFT.

Additionally, non-dilute $H_i$ structures were generated with between 2 and 32 $H_i$, corresponding to $y = 0.125$ to 2 in $H_yLa_{0.5}Sr_{0.5}CoO_{3-\delta}$. For non-dilute defects, the number of unique configurations quickly exceeds what is computationally feasible to comprehensively evaluate, so we randomly sampled up to 1,000 unique configurations for each $H_yLa_{0.5}Sr_{0.5}CoO_{3-\delta}$ composition ($y$ ranging from 0.125 to 2 and $\delta = 0, 0.25, 0.5$). These configurations were first relaxed using UMA. The resulting lowest energy UMA structure at each composition was relaxed using DFT. To gain additional insights into the influence of $H_i$ formation on $v_O$ formation, we also enumerated all unique single $v_O$ for these lowest energy $H_i$-containing structures. These were relaxed using UMA. Taken together, our calculations span >100 unique compositions and >50,000 unique configurations evaluated across the $H_yLa_{0.5}Sr_{0.5}CoO_{3-\delta}$ composition space.

## III. Results and Discussion

Although prior experimental and computational efforts have extensively investigated the protonated BM SrCoO$_{2.5}$ phase, no study has thoroughly investigated the energetics of protonation across the P-to-BM transition in mixed La/Sr cobaltites. [24,32,34,38–40] Motivated by recent reports of protonated LSCO phases [41,42] and degradation of LSCO in electrolyte-gated devices by acid-etching [12,20,26], we examined the thermodynamics of protonation across the P-to-BM transition for La$_{0.5}$Sr$_{0.5}$CoO$_{3-\delta}$.

### A. Plausibility of protonation

To quantify the tendency for $v_O$ or $H_i$ formation through the P-to-BM transition, dilute defect formation energies were calculated using DFT according to Equation 1. The lowest dilute defect formation energies for each defect type are shown for $p_{O_2}$ and $p_{H_2}$ of 1 bar at 300 K across the P-to-BM transition in **Figure 2**, where the darker bars indicate the dilute $v_O$ formation energy and the lighter bars indicate the dilute $H_i$ formation energy. The P phase $v_O$ formation energy of 1.08 eV agrees well with prior values in literature. [82–85] Moving across the transition, the $v_O$ formation energy increases slightly to 1.37 eV for the $\delta = 0.25$ phase, while increasing drastically to 3.04 eV for BM. The ~2 eV increase in $v_O$ formation energy across the transition is indicative of the stability of BM against further $v_O$ formation. The stability of BM compared to P can be partially attributed to strain minimization, with the regularly ordered alternating CoO$_6$ octahedra and CoO$_4$ tetrahedral layers in BM minimizing local strain that is otherwise present with disordered defects. However, the oxidation state of Co also plays a critical role. For P La$_{0.5}$Sr$_{0.5}$CoO$_3$, the Co oxidation state is nominally composed of a mix of 3+ and 4+ states, with an average value of 3.5+. The Co$^{4+}$ state is generally less stable than Co$^{3+}$ or Co$^{2+}$, rationalizing why SrCoO$_3$ (nominally Co$^{4+}$) has a lower $v_O$ formation energy than LaCoO$_3$ (nominally Co$^{3+}$) [82,85] and P LSCO a lower $v_O$ formation energy than BM LSCO.

$H_i$ formation energies present a different picture than the $v_O$ formation energies. In the P phase, the dilute $H_i$ formation energy is negative, and while it increases slightly with increasing δ, it remains negative even in the BM phase. This suggests facile insertion of hydrogen across the P-to-BM transition at the given conditions ($p_{H_2}$ = 1 bar, 300 K). As we show later, and in agreement with previous works on BM SrCoO$_{2.5}$ [38,40], the insertion of hydrogen in LSCO occurs through the formation of hydroxyl (-OH) bonds which reduce nearby Co. Unlike the case of $v_O$ formation, this apparent reduction does not become significantly less favorable as the host structure becomes more reduced (moving from δ = 0 to δ = 0.5).

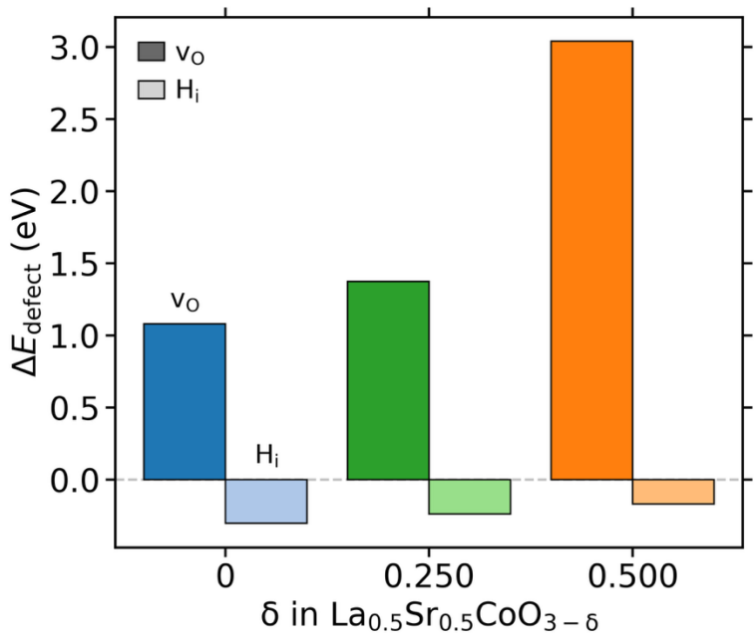

**Figure 2.** DFT-calculated lowest-energy dilute oxygen vacancy ($v_O$) and dilute hydrogen interstitial ($H_i$) formation energies ($\Delta E_{\text{defect}}$) across the topotactic transition of La$_{0.5}$Sr$_{0.5}$CoO$_{3-\delta}$ from perovskite (δ = 0) to brownmillerite (δ = 0.5). Defect formation energies are calculated assuming chemical potentials consistent with $p_{O_2}$ and $p_{H_2}$ of 1 bar at 300 K. Dark bars correspond to dilute $v_O$, while light bars correspond to dilute $H_i$. Blue denotes the perovskite (δ = 0) host structure, green the δ = 0.25 host structure, and orange the brownmillerite (δ = 0.5) host structure.

For the conditions denoted in **Figure 2**, the dilute defect formation energies suggest that $H_i$ formation should be thermodynamically favorable and strongly preferred relative to $v_O$ formation across the P-to-BM transition. However, these reference conditions may not be consistent with the operation of most devices. The effect of changing environmental conditions on defect formation energies can be understood through shifts in the oxygen and hydrogen chemical potentials as captured in Equations 4 and 6 and applied in Equation 1. Elevated temperatures and lower $p_{O_2}$ decrease $\mu_O$, lowering the $v_O$ formation energy. Higher temperatures and lower $p_{H_2}$ also decrease $\mu_H$, but this increases the $H_i$ formation energy since hydrogen atoms are removed from the environment. In the electrochemical setting, an applied positive voltage lowers both the $v_O$ and $H_i$ formation energies, though $v_O$ is lowered doubly compared to $H_i$ due to the charge of an $O^{2-}$ ion being twice that of $H^+$. To better understand the competition between $v_O$ and $H_i$ formation, the difference between their defect formation energies was evaluated across a set of conditions as presented in **Figures 3** and **S2**.

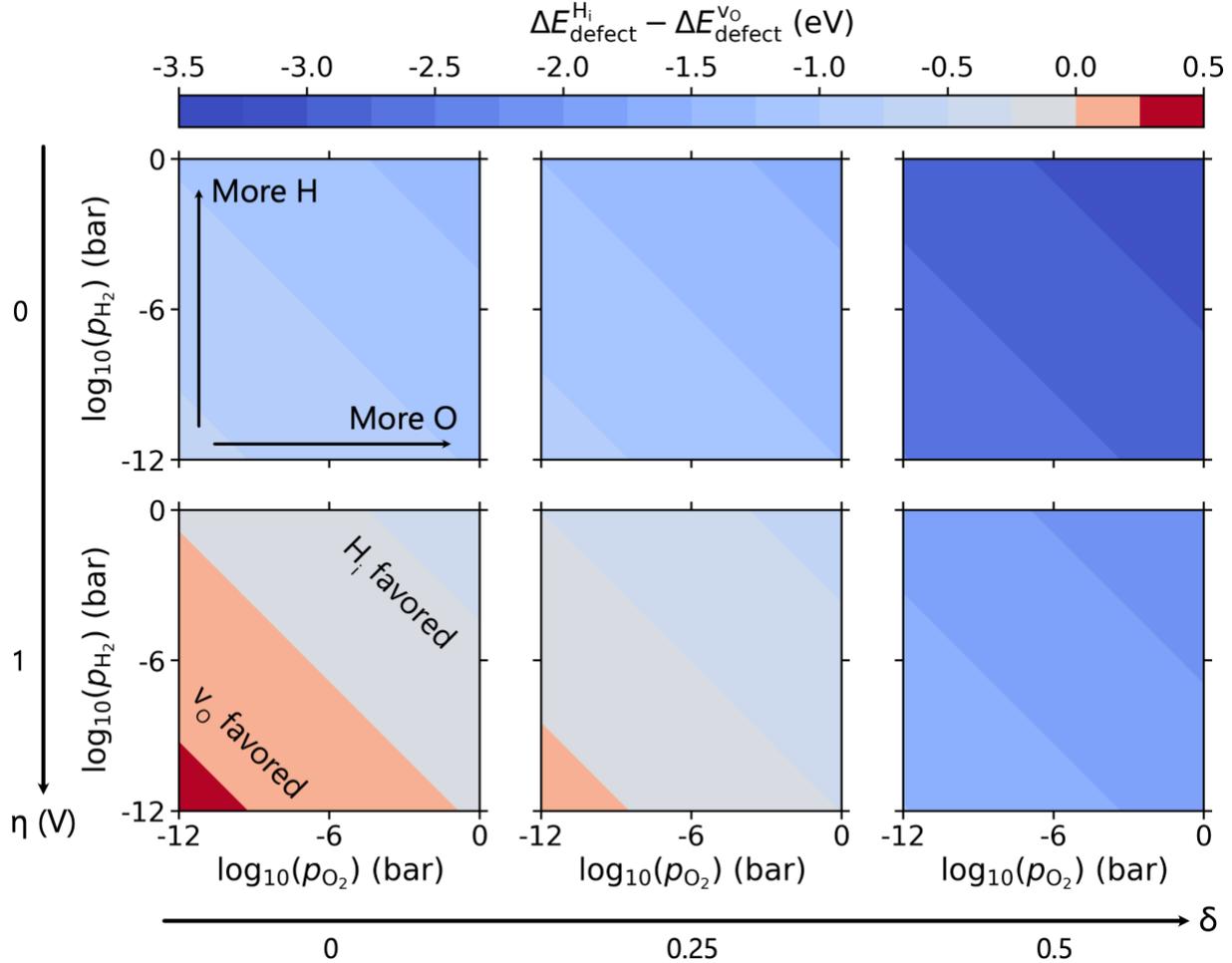

**Figure 3.** Defect competition between dilute oxygen vacancy ($v_O$) and dilute hydrogen interstitial ($H_i$) formation across the topotactic transition of $La_{0.5}Sr_{0.5}CoO_{3-\delta}$ from perovskite ($\delta = 0$) to brownmillerite ($\delta = 0.5$). Dilute formation energies ($\Delta E_{defect}$) were determined using DFT. Blue shading indicates conditions where dilute $H_i$ formation has a lower energy than dilute $v_O$ formation, whereas red indicates the opposite. All dilute defect formation energies were calculated at 300 K. The first row corresponds to no applied electrochemical potential ($\eta$), while the second row corresponds to an applied potential of 1 V. The lower left of each subplot reflects ultra-high vacuum (or ultra-low concentration) conditions for oxygen and hydrogen, while the upper right reflects higher partial pressure (concentration) conditions up to 1 bar. Analogous calculations performed at 1000 K are presented in **Figure S1**.

In **Figure 3**, the first row shows the thermodynamic competition between $v_O$ and $H_i$ formation as a function of partial pressures (concentrations) at 300 K and no applied electrochemical potential, with the subplots from left to right corresponding to the P ($\delta = 0$), $\delta = 0.25$, and BM ($\delta = 0.5$) hosts. The second row shows the competition over the same range of conditions with 1 V of applied potential. Blue shading indicates where $H_i$ formation is thermodynamically favored over $v_O$ formation, while red shading shows where $v_O$ formation is favored. In the absence of an applied potential, $H_i$ formation is more favorable than $v_O$ formation at 300 K for all three host structures, across all concentrations evaluated. Under an applied potential of 1 V, shown in the second row of **Figure 3**, the doubly charged nature of $O^{2-}$ shifts the thermodynamic competition. In the P phase, $v_O$ formation is now favored under vacuum-like (low concentration) conditions. $H_i$ formation remains more favorable than $v_O$ formation for the more reduced hosts ($\delta = 0.25$ and BM), though the preference is weakened. While these results generally suggest a tendency for hydrogen insertion at equilibrium, it is important to note that $H_i$ formation may be more sensitive to the system setup as hydrogen must come in contact with the LSCO surface to initiate insertion. In contrast, $v_O$ formation can proceed simply by liberating $O_2$ from LSCO. In the context of electrolyte-gated devices, the low-concentration regime where $v_O$ formation is preferred under an applied potential may be indicative of vacuum-like operation. For device operation under higher relative humidity, residual $H_2O$ in the electrolyte could serve to increase the hydrogen concentration by voltage-induced water-splitting. These calculations suggest that hydrogen should insert into LSCO at various stages of reduction if hydrogen (and oxygen) are available in non-negligible concentrations. Decreasing hydrogen and oxygen concentrations and/or an increasingly positive applied bias encourages $v_O$ rather than $H_i$ formation.

The effect of temperature is explored in **Figure S1**, where the competition between dilute $v_O$ and $H_i$ formation is shown for 300 K and 1000 K, with no applied potential. The partial pressure effects on chemical potentials become much more prominent at 1000 K, leading to lower $v_O$ and higher $H_i$ formation energies. This results in $v_O$ formation being favored across nearly all conditions for the more oxidized host structures (P and $\delta = 0.25$). Once the BM phase is reached, $v_O$ formation is slightly favored for high vacuum conditions, while $H_i$ formation is favored at higher pressures (concentrations). These findings suggest that $H_i$ formation is not an impediment to the high-temperature synthesis of oxygen-deficient LSCO or the use of LSCO in high-temperature applications such as thermochemical water-splitting or as an oxide-ion conductor for solid oxide fuel cells. For the latter cases, this proves especially true for the more oxidized states (e.g., P). This observation qualitatively aligns with the prior report where upon heating to 175 °C, $HSrCoO_{2.5}$ transforms into $SrCoO_{2.5}$. [24]

Although our calculations only consider $x = 0.5$, we can briefly speculate on implications for other $x$ in $La_{1-x}Sr_xCoO_{3-\delta}$. For higher $x$, $v_O$ formation energies are known to be lower due to the instability of the $Co^{4+}$ oxidation state. [82,85] Our results in **Figure 2** suggest that dilute $H_i$ formation energies are less sensitive to the Co valence. This implies that $v_O$ formation may be favorable (more competitive with $H_i$ formation) across a broader range of conditions for compositions with more Sr. This could explain in part why significant $H_i$ formation is not observed in $SrCoO_{3-\delta}$ until the BM phase is reached [24], while there is a report of an $H_{0.67}La_{0.67}Sr_{0.33}CoO_{2.67}$ phase [41] for $x = 0.33$.

**B. Limits on protonation**

Much as the $v_O$ formation energy increases significantly across the P-to-BM transition, there should be some concentration of $H_i$ above which further protonation of LSCO becomes unfavorable. To investigate this limit, non-dilute $H_i$ ($\Delta n_H \geq 1$) configurations were investigated for each host structure ($\delta$ = 0, 0.25, 0.5).

The growing number of possible configurations at higher $H_i$ concentrations motivates the use of uMLIPs. The applicability of the UMA uMLIP [60] for evaluating defective LSCO configurations was assessed by comparing the energies of a set of host and dilute defect structures from DFT against those obtained from UMA relaxations, shown in **Figure S2**. The DFT and UMA total energies are directly compared via a parity plot in **Figure S2a**, where the UMA energies are consistently offset from the parity line. This may be due to UMA being trained on GGA+U calculations with an effective Hubbard value of 3.32 eV for Co instead of the 3.0 eV used in this work. [60] Fitting a linear correction to remove systematic error results in a mean absolute error of ~4 meV/atom for UMA predictions of DFT total energies, with no significant outliers, shown in **Figure S2b**. UMA also identifies the same ground-state configurations as DFT. The applicability of UMA is further supported by a comparison of DFT forces against UMA predictions of forces for the DFT-relaxed geometries, with the distribution of errors plotted in **Figure S3**. The distribution is centered tightly at zero for both host and dilute defect structures with an overall absolute error of 0.093 eV/Å. The low error in energies, identification of the correct ground-state configurations, and low error in forces demonstrate that UMA is reasonably well-suited for the study of the HLSCO system without fine-tuning. Importantly, no systematic correction to the energy was applied in practice in this work, as the proposed linear correction would cancel for the evaluation of defect formation energies (Equation 1).

To evaluate non-dilute $H_i$ energetics, many configurations were evaluated at each unique composition ($0.0625 \leq y \leq 2$ and $\delta$ = 0, 0.25, 0.5 in $H_yLa_{0.5}Sr_{0.5}CoO_{3-\delta}$), using UMA to fully relax the atomic positions and lattice parameters. The same configurations were additionally relaxed with fixed cell conditions, where the atom positions could relax, but the lattice parameters remained fixed. This was done to isolate the effect of $H_i$-induced lattice strain from Co reduction when interpreting trends in the $H_i$ formation energy with changing $y$. The UMA-predicted $H_i$ defect formation energies for the lowest energy defective structures (where $\Delta n_H$ = 1 to 32) are shown in **Figure 4**, along with the corresponding interaction energies determined according to Equation 7 (with $n_{v_O} = 0$), and the associated change in the pseudo-cubic lattice parameter. Defect formation energies and interaction energies are shown on a per formula unit basis ($y\Delta E_{\text{defect}}^{H_i}$, $y\Delta E_{int}$) to maintain a consistent scale.

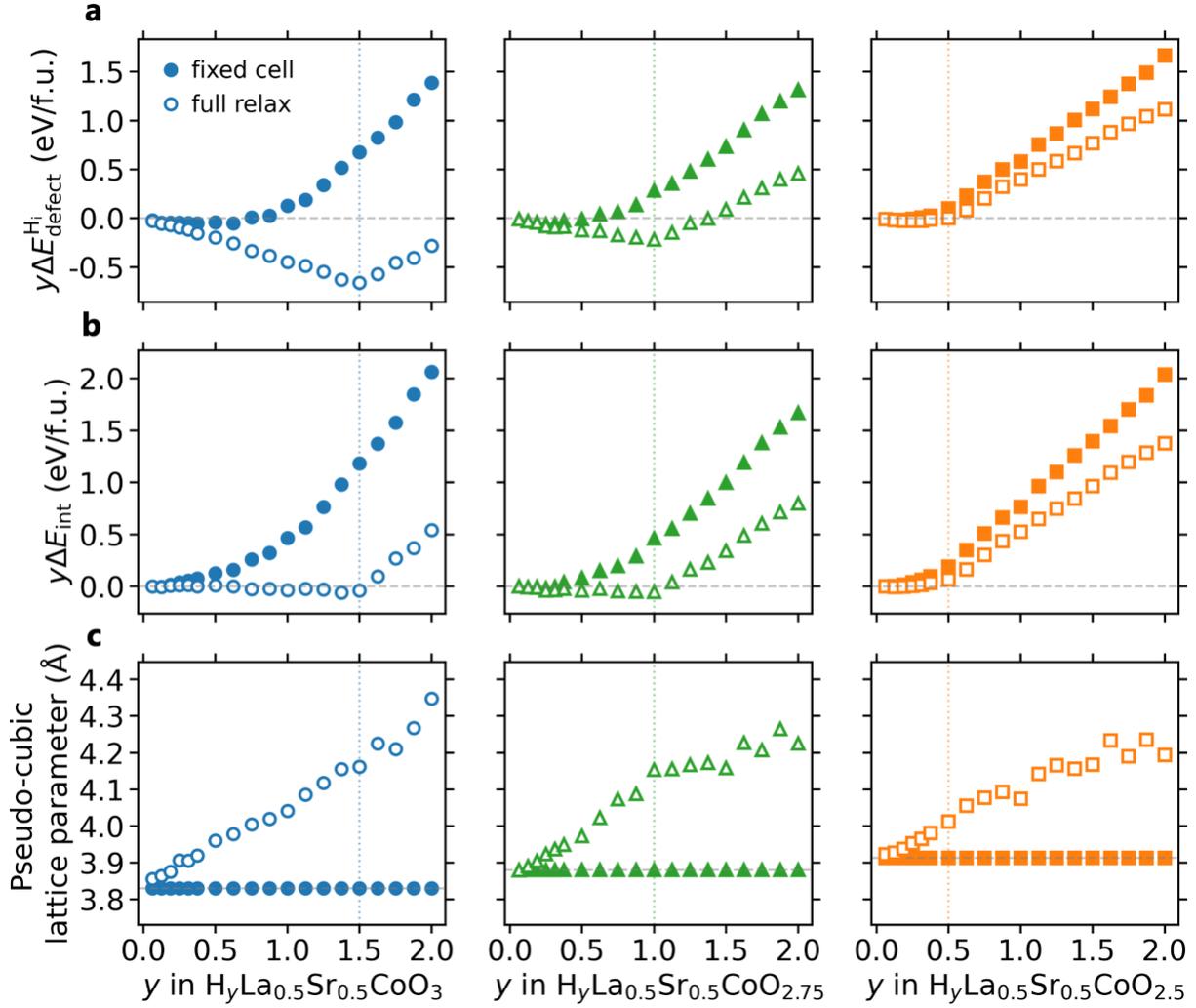

**Figure 4.** UMA [60] calculations of non-dilute hydrogen interstitials ($H_i$) across the topotactic transition of $La_{0.5}Sr_{0.5}CoO_{3-\delta}$ from perovskite ($\delta = 0$) to brownmillerite ($\delta = 0.5$). **a)** $H_i$ formation energies ($y\Delta E_{defect}^{H_i}$) relative to the pristine (no $H_i$) host for varying concentrations of $H_i$. Horizontal grey dashed lines correspond to defect formation energies of 0 eV/f.u. **b)** Interaction energies ($y\Delta E_{int}$) between $H_i$ for varying concentrations of $H_i$. Horizontal dashed lines correspond to defect interaction energies of 0 eV/f.u. **c)** Pseudo-cubic lattice parameter tracking structural expansion with increasing concentration of $H_i$. Horizontal grey dashed lines correspond to the pseudo-cubic lattice parameter of the pristine host. For all calculations, the chemical potential conditions are consistent with $p_{H_2}$ of 1 bar at 300 K. In these plots, $\Delta E_{defect}^{H_i}$ and $\Delta E_{int}$ are on a per defect basis. Multiplying them by $y$ (the molar composition of hydrogen per formula unit) converts them to a per formula unit basis to maintain a consistent scale. Across the subplots, vertical dotted lines correspond to compositions where Co nominally has an oxidation state of 2+. Blue circles in the first column pertain to the perovskite ($\delta = 0$) host structure, green triangles in the second column the $\delta = 0.25$ host structure, and orange squares in the third column the brownmillerite ($\delta =$

0.5) host structure. Open markers denote calculations where both atom positions and lattice parameters were fully relaxed. Closed markers correspond to calculations where atoms positions were relaxed, but lattice parameters remained fixed.

For the fully relaxed structures, the $H_i$ formation energy initially decreases with increasing $y$ for each host. Notably, the per defect formation energies are often more negative than the dilute $H_i$ formation energy. For example, at $y = 1$ and $\delta = 0$ (**Figure 4a**), the per defect formation energy of -0.45 eV/$H_i$ is 0.15 eV lower than the dilute formation energy in the same host (-0.30 eV, **Figure 2**). DFT calculations were also performed on this UMA-predicted ground-state structure and yield a per defect energy of -0.44 eV/$H_i$ for $y = 1$, supporting the accuracy of UMA for these higher levels of proton insertion.

Examining the interaction energies across all three hosts (**Figure 4b**), they remain near zero up to $y = 1.5$ for P, $y = 1$ for $\delta = 0.25$, and $y = 0.5$ for BM (denoted by vertical dotted lines in **Figure 4**) before increasing sharply. This trend in interaction energy gives rise to the observed minima in the $H_i$ formation energy (**Figure 4a**). The compositions where these minima occur all nominally correspond to $Co^{2+}$, the lowest stable oxidation state of Co. Further protonation would require a different redox mechanism to maintain charge neutrality in the material. Experimentally, it has been suggested that hydrogen insertion beyond the reduction limit is possible in BM $SrCoO_{2.5}$ via the formation of $H_2$ dimers. [32] Investigating the $y = 2$ structures reveals UMA-predicted $H_2$ dimer formation across all three hosts, in agreement with these findings.

Examining the pseudo-cubic lattice parameter shows that for each host, increasing $H_i$ concentration leads to significant expansion of the structure. In particular, comparing the $y = 0$ baselines across hosts with the expansion caused by $H_i$ formation in **Figure 4c**, achieving a protonation level of $y = 0.5$ in the P phase leads to approximately the same expansion as the full P-to-BM transition according to our calculations. The role of volume expansion in stabilizing $H_yLa_{0.5}Sr_{0.5}CoO_{3-\delta}$ structures becomes clear when considering the "fixed cell" relaxations. When the cell is constrained, local strain-related energy penalties become much more pronounced, leading to more positive $H_i$ formation energies at lower $H_i$ concentrations. Notably though, the $H_i$ formation energy remains near or below zero for a meaningful range of $H_i$ concentrations. In addition to full cell and fixed cell relaxations, UMA calculations were carried out only allowing c-axis expansion, mirroring the case of epitaxial films. The resulting $H_i$ formation energies follow those of the fully relaxed case very closely, being slightly higher in energy, while having the minimum energy still occur at the Co reduction limit. These results demonstrate that as long as there is Co redox availability, LSCO is able to accommodate $H_i$ formation across the P-to-BM transition.

In addition to the LSCO phases considered in this work, protonated BM $SrCoO_{2.5}$ (SCO) configurations were examined with UMA to further validate these findings. The results are shown in **Figure S4**, with a predicted minimum $H_i$ formation energy near $HSrCoO_{2.5}$ ($y = 1$), the same composition commonly suggested to be formed experimentally in electrolyte-gated devices and which nominally corresponds to $Co^{2+}$. [24,32,34] Prior DFT-based computational works have also found the same trends for $H_ySrCoO_{2.5}$. [40] Furthermore, DFT relaxation of the UMA-predicted ground-state structures for the $H_yLa_{0.5}Sr_{0.5}CoO_{3-\delta}$ compositions are shown to closely reproduce the UMA results (**Figure S5**). Given our findings for $H_yLa_{0.5}Sr_{0.5}CoO_{3-\delta}$ and $H_ySrCoO_{2.5}$, it is reasonable to project similar behavior for other La:Sr ratios, where the $H_i$ insertion limit (without dimer formation) is tied to the Co reduction limit ($Co^{2+}$).

Previously, it has been suggested that the presence of $H_i$ may promote $v_O$ formation (i.e., the interaction energy between $H_i$ and $v_O$ is negative). [42] To explore this effect, we used UMA to calculate all unique $v_O$ formation energies for each of the UMA-identified lowest energy $H_i$ configurations, treating the $H_i$-only structures as the host when calculating the $v_O$ formation energy. The dilute $v_O$ formation energies as calculated from UMA vary slightly from DFT, as shown in **Table S1**. To facilitate comparison with results in the previous section, all UMA-predicted $v_O$ formation energies were shifted by the difference in these values to reproduce the DFT dilute $v_O$ formation energies at $y = 0$. The resulting $v_O$ formation and interaction energies are shown in **Figure 5**.

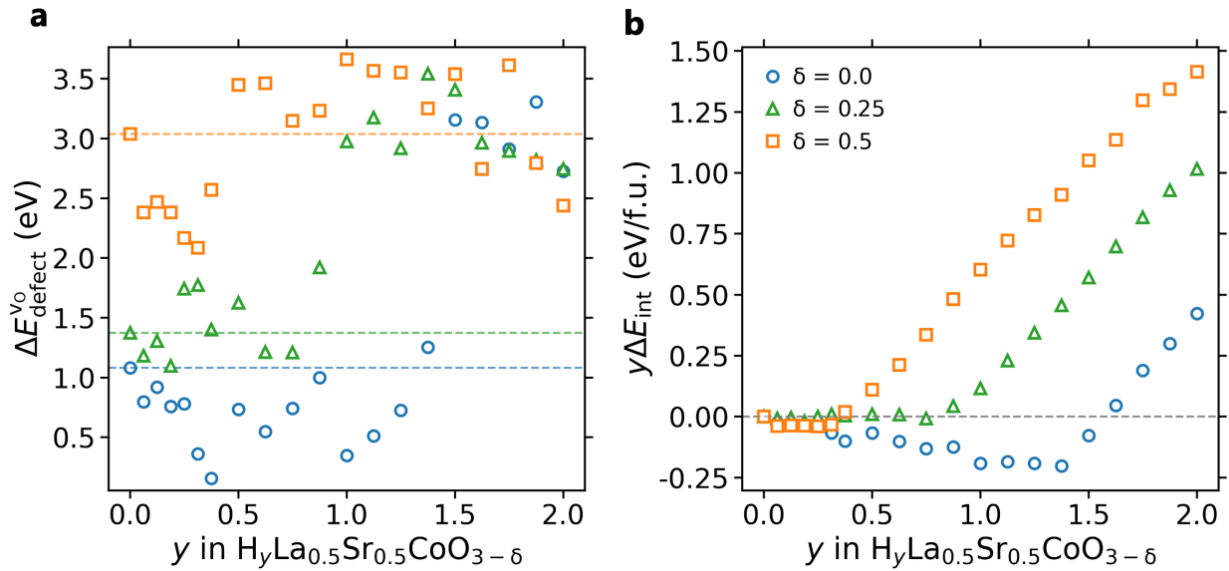

**Figure 5. a)** UMA-calculated dilute oxygen vacancy ($v_O$) formation energy as a function of hydrogen interstitial ($H_i$) concentration ($y$) across the topotactic transition of $La_{0.5}Sr_{0.5}CoO_{3-\delta}$ from perovskite ($\delta = 0$) to brownmillerite ($\delta = 0.5$). Dilute $v_O$ formation energies ($\Delta E_{defect}^{v_O}$) were calculated treating the protonated structure as the host. All UMA-derived $v_O$ formation energies were shifted to match the DFT-derived energies for the $y = 0$ host case (see **Table S1**). The dashed horizontal lines denote the dilute $v_O$ formation energies in the absence of $H_i$. **b)** Defect interaction energy ($y\Delta E_{int}$) between the single $v_O$ and multiple $H_i$. $\Delta E_{int}$ is on a per defect basis where multiplying by $y$ converts to a per formula unit basis to maintain a consistent scale with **Figure 4**. Chemical potential conditions are consistent with $p_{O_2}$ and $p_{H_2}$ of 1 bar at 300 K. In both subplots,

blue circles correspond to the perovskite ($\delta = 0$) host structure, green triangles the $\delta = 0.25$ host structure, and orange squares the brownmillerite ($\delta = 0.5$) host structure.

Below the reduction limit of Co, protonation is found to lower the $v_O$ formation energy by up to ~1 eV, though the strength of this effect is strongly dependent on the particular composition and configuration. Meanwhile, at and above the reduction limit, $v_O$ formation energies increase to roughly those found for BM (see **Figure 2**). Accordingly, the interaction energies are also negative below the Co reduction limit, confirming that $H_i$ can promote $v_O$ formation. Although $v_O$ formation never becomes spontaneous ($\Delta E_{\text{defect}}^{v_O} < 0$), this has possible implications for the previous defect competition analysis (**Figure 3**), which only examined the competition between dilute $v_O$ and $H_i$ defects across the P-to-BM transition. For both $v_O$ and $H_i$, non-dilute $H_i$ concentrations are found to lower subsequent defect formation. However, in the case of $v_O$, especially for P, the interaction energy is slightly more negative. As $H_i$ accumulate in LSCO, this suggests that the $v_O$ would become comparatively more favorable over a wider range of conditions.

## C. Thermodynamic stability

Thus far, the defect thermodynamics of $v_O$ and $H_i$ formation across the P-to-BM transition in LSCO suggests facile insertion of H is possible up to the Co reduction limit. To understand the extent to which these protonated structures are stable with respect to phase separation, a convex hull analysis was carried out. A first analysis was conducted for the phase space encompassing the "topotactic" transition, where it is assumed that a perovskite-like structure is conserved. Next, stability with respect to the entire H-La-Sr-Co-O phase space was examined. All structures in this chemical space that are reported to be thermodynamically stable in Materials Project were relaxed with UMA to yield compatible energies with the UMA-relaxed structures generated in this work. [63,75,76]

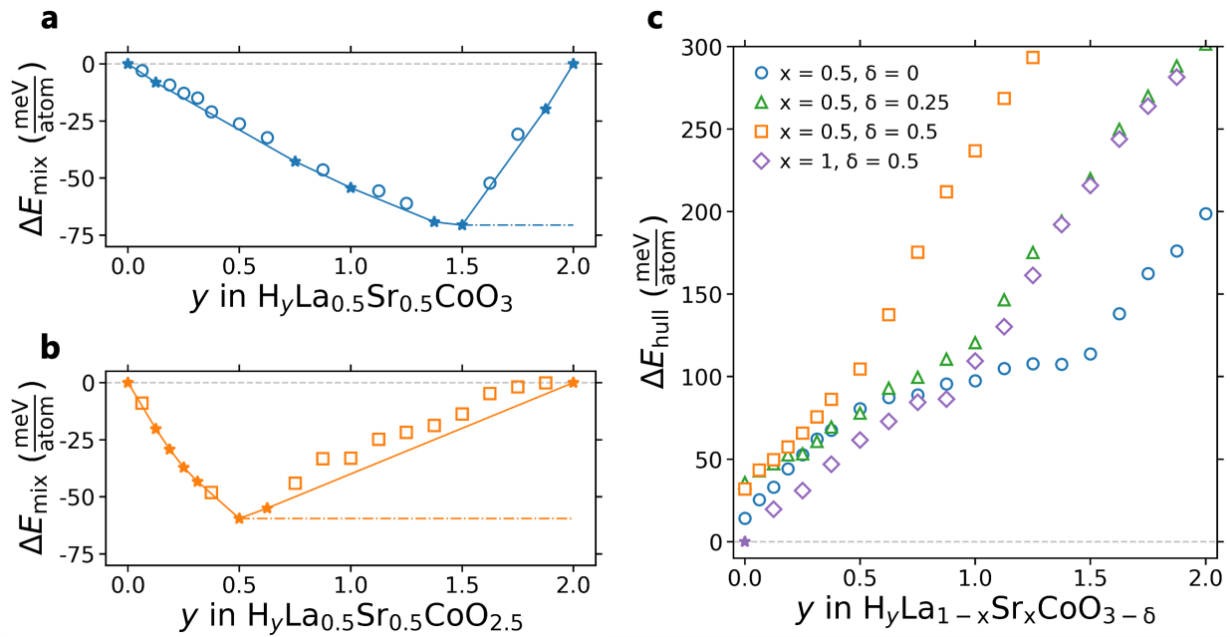

**Figure 6. a)** Mixing energies ($\Delta E_{mix}$) relative to $La_{0.5}Sr_{0.5}CoO_3$ and $H_2La_{0.5}Sr_{0.5}CoO_3$ where the solid blue line denotes the mixing hull, and stars denote phases that are stable against decomposition relative to these two endpoints. The dashed-dotted line represents the energy of $H_{1.5}La_{0.5}Sr_{0.5}CoO_3$ + 0.25 $H_2$. **b)** Mixing energies ($\Delta E_{mix}$) relative to $La_{0.5}Sr_{0.5}CoO_{2.5}$ and $H_2La_{0.5}Sr_{0.5}CoO_{2.5}$ where the solid orange line denotes the mixing hull, and stars denote phases that are stable against decomposition relative to these two endpoints. The dashed-dotted line represents the energy of $H_{0.5}La_{0.5}Sr_{0.5}CoO_{2.5}$ + 0.75 $H_2$. **c)** Energy above the hull ($\Delta E_{hull}$) for $H_yLa_{0.5}Sr_{0.5}CoO_{3-\delta}$ and $H_ySrCoO_{2.5}$ phases when considering a convex hull built from these phases and all hull stable compounds in the H-La-Sr-Co-O chemical space queried from Materials Project [75,76]. Stars denote stable phases on this "complete" hull. Blue circles correspond to the $x = 0.5$ perovskite ($\delta = 0$) host structure, green triangles the $x = 0.5$ and $\delta = 0.25$ host structure, orange squares the $x = 0.5$ brownmillerite ($\delta = 0.5$) host structure, and purple diamonds the $x = 1$ brownmillerite ($\delta = 0.5$) host structure for $La_{1-x}Sr_xCoO_{3-\delta}$. Energies for all phases were determined with UMA [60]. All calculations are performed assuming a temperature of 0 K.

In the topotactic convex hulls shown in **Figures 6a** and **6b**, we start from P ($\delta = 0$) or BM ($\delta = 0.5$), presume that the $y = 0$ ($La_{0.5}Sr_{0.5}CoO_{3-\delta}$) and $y = 2$ ($H_2La_{0.5}Sr_{0.5}CoO_{3-\delta}$) phases are stable end-members, and consider the tendency to mix these phases. In both P and BM, the most negative mixing energy can be found at the Co reduction limit ($y = 1.5$ for P, $y = 0.5$ for BM). At lower hydrogen insertion compositions (smaller $y$), many compositions are found to lie just barely on the convex hull. This suggests the possibility of solid solutions between the host oxide ($y = 0$) and the protonated phases at the Co reduction limit at finite temperatures. The dashed-dotted lines represent the mixing energy at higher $y$ if HLSCO is allowed to exchange hydrogen with its environment (e.g., the energy of $H_{1.5}La_{0.5}Sr_{0.5}O_3$ + 0.25 $H_2$ at $y = 2$ for $\delta = 0$). Under this scenario, any protonated phases beyond the Co reduction limit quickly become unstable.

Analysis of hull stability in the full H-La-Sr-Co-O space reveals a different picture. As a baseline, the stability of BM $SrCoO_{2.5}$, which has been assessed in previous works via DFT [40], was calculated to validate the applicability of UMA to hull stability calculations. The stability of BM SCO as a function of $y$ is shown by the purple circles in **Figure 6c**. UMA predicts the host BM phase $SrCoO_{2.5}$ to be stable, while the protonated composition $H_{0.125}SrCoO_{2.5}$ is slightly unstable with $\Delta E_{hull} = 20$ meV/atom. This is comparable to the DFT results from literature which predict the BM phase of $SrCoO_{2.5}$ to be stable and $H_{0.125}SrCoO_{2.5}$ unstable with $\Delta E_{hull} = 19$ meV/atom. [40]

All $H_yLa_{0.5}Sr_{0.5}CoO_{3-\delta}$ phases are found to have $\Delta E_{hull} > 0$, indicating they are unstable with respect to decomposition (**Figure 6c**). This includes the host structures, among which P ($y = 0$) lies closest to the convex hull. This agrees with prior experimental work suggesting La occupation of the *A*-site increases the stability of the P phase with respect to BM. [15] Across the P-to-BM transition, hydrogen insertion leads to increasing $\Delta E_{hull}$. Notably, up to $y = 1$, all LSCO compositions are more unstable than BM SCO, though all protonated phases of BM SCO are also found to be unstable. The most common decomposition products include $Co_3O_4$, $Sr(OH)_2$, $Co_3H$, $Sr_6Co_5O_{15}$, $La_2O_3$, and LaOOH. In the context of the highly favorable defect formation energies, these results are surprising. They suggest that although hydrogen insertion is thermodynamically favorable from a defect formation perspective, the obtained protonated phases are metastable with respect to decomposition by phase separation. This offers a possible explanation for the acid-etching effect observed in EDLT experiments involving LSCO. In this scenario, $H^+$ may be generated from $H_2O$ via electrolysis at positive gating-voltages, driven towards and across the electrolyte-LSCO interface, forming hydrogen interstitials. Once incorporated, the driving force for decomposition in the protonated LSCO phase could be sufficiently large to trigger breakdown of the host P or BM structure. If this degradation mechanism is at play, the lower $\Delta E_{\text{hull}}$ for SCO suggests that the relative stability with respect to decomposition can be tuned via the *A*-site composition. [86]

**D. Consequences of protonation**

The thermodynamic analysis of protonated LSCO informs: 1) that protonation of LSCO is thermodynamically favorable and competitive with $v_O$ formation across a range of conditions and 2) that protonated LSCO phases are unstable with respect to decomposition into mixtures of phases. Although they are thermodynamically unstable with respect to the full H-La-Sr-Co-O phase diagram, they are still likely to appear as metastable phases during the operation of electrochemical devices like EDLTs since the rates of ion insertion and removal should be much higher than the rates of phase separation. Previous reports of BM $HSrCoO_{2.5}$, which is calculated to be unstable relative to the H-Sr-Co-O phase diagram, support the metastability of these phases. [24] This section explores the consequences of protonation when a perovskite-like structure is preserved.

The limited reports of experimentally studied $H_yLa_{1-x}Sr_xCoO_{3-\delta}$ phases ($HSrCoO_{2.5}$, $H_{0.67}La_{0.67}Sr_{0.33}CoO_{2.67}$, and $H_yLa_{0.7}Sr_{0.3}CoO_{3-\delta}$) show increasing resistivity and widening of the band gap upon protonation. [24,32,41,42] We assessed this effect by analyzing the band gap from our DFT calculations of the UMA-identified lowest energy structures across three magnetic initializations (FM, G-AFM, and A-AFM). These magnetic states are energetically close to one another for many of the phases, as shown in **Figure S6**, but can have a strong effect on the observed band gap. Previous work on BM SCO also noted that the choice of Hubbard U value and lattice constants can affect which magnetic configuration is found to be the ground-state via GGA+U calculations. [52] To understand the general trend, the average band gap found from all three magnetic initializations is shown as a function of *y* for the three host structures across the P-to-BM transition in **Figure 7a**. The P host retains a near-zero band gap for low levels of protonation, before widening near-monotonically thereafter. For $\delta = 0.25$ and BM, protonation initially has the effect of slightly shrinking or widening the band gap, before increasing rapidly at higher *y*. In all three cases, the band gap at the Co reduction limit is ~1.5 eV, significantly higher than the ~0 eV and ~0.5 eV averages found for unprotonated P and BM hosts, respectively.

To illustrate the effect of protonation on the electronic structure, the density of states (DOS) for the FM-initialized calculations of P $La_{0.5}Sr_{0.5}CoO_3$ and $H_{1.5}La_{0.5}Sr_{0.5}CoO_3$ are shown in **Figure 7c**, while the G-AFM-initialized calculations of BM $La_{0.5}Sr_{0.5}CoO_{2.5}$ and $H_{0.5}La_{0.5}Sr_{0.5}CoO_3$ are shown in **Figure 7d**. The unprotonated P and BM phases exhibit band gaps of 0 and 0.72 eV, respectively, in reasonable agreement with experiment. [15,87] Upon protonation to their Co reduction limit, the band gap in both widens to ~1.5 eV, in alignment with the general trend observed previously. Notably, for all four DOS, O 2$p$ and Co 3$d$ states compose the majority of the valence and conduction bands near the Fermi level. Comparing the unprotonated versus protonated phases, protonation shifts the occupied O 2$p$ band center down and unoccupied 3$d$ band center up relative to the Fermi level. Considering that inserted hydrogen forms -OH bonds while reducing nearby Co, protonation necessarily requires electron transfer between these bands. The increasing gap between these band centers should make partially protonated LSCO less prone to further protonation. [88]

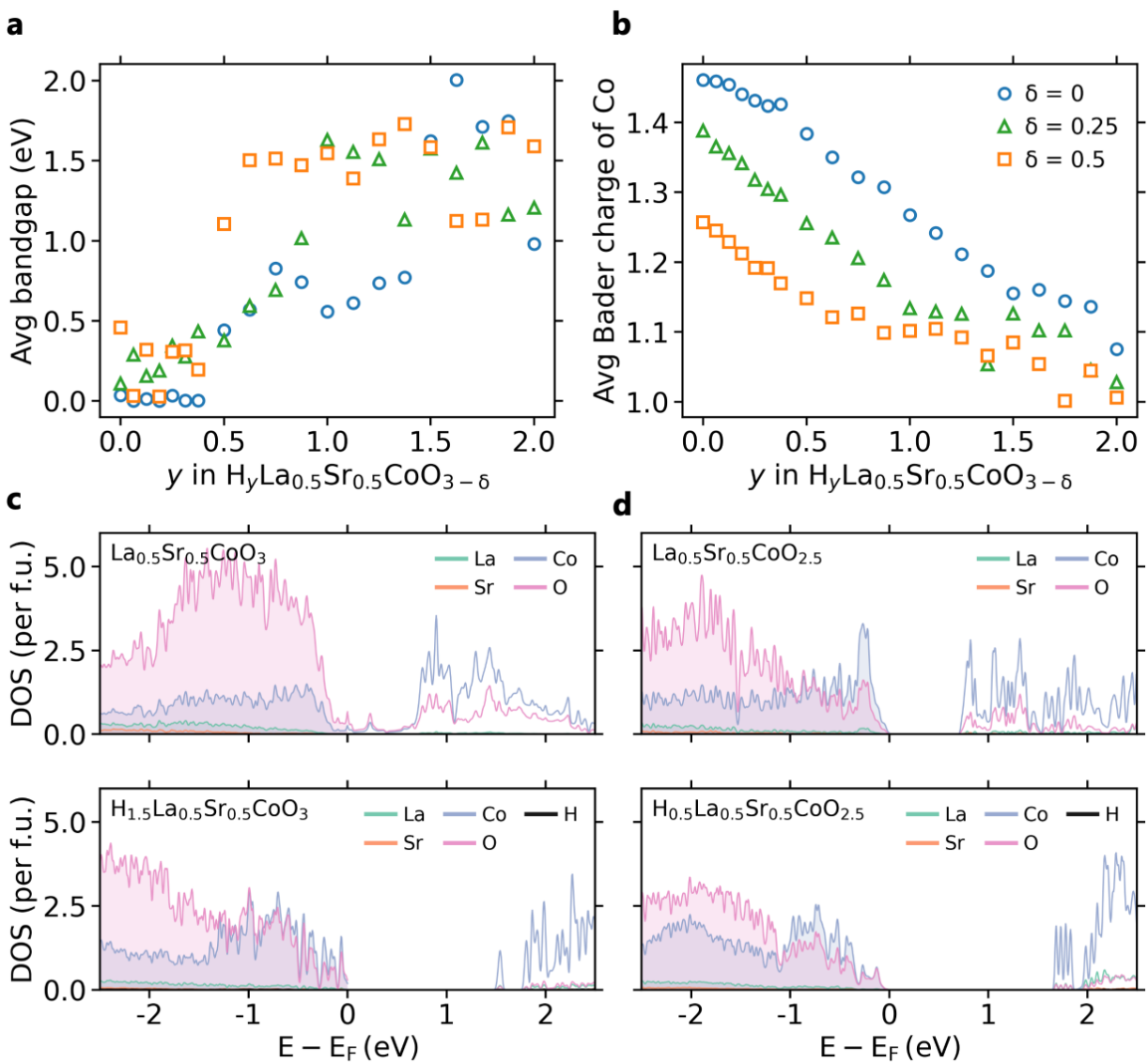

**Figure 7.** DFT calculations of **a)** average band gap and **b)** average Co Bader charge across three magnetic moment initializations (FM, G-AFM, A-AFM) of UMA [60]-identified ground-state structures for $y = 0$ to $y = 2$ in $H_yLa_{0.5}Sr_{0.5}CoO_{3-\delta}$ ($\delta = 0, 0.25, 0.5$). Blue circles correspond to the perovskite ($\delta = 0$) host structure, green triangles the $\delta = 0.25$ host structure, and orange squares the brownmillerite ($\delta = 0.5$) host structure. **c)** Element-projected density of states (DOS) for FM-initialized $La_{0.5}Sr_{0.5}CoO_3$ and $H_{1.5}La_{0.5}Sr_{0.5}CoO_3$ ($\delta = 0$). **d)** Element-projected DOS for G-AFM-initialized $La_{0.5}Sr_{0.5}CoO_{2.5}$ and $H_{0.5}La_{0.5}Sr_{0.5}CoO_{2.5}$ ($\delta = 0.5$). The DOS are normalized per formula unit and the energy is presented relative to the Fermi energy ($E_F$) with elements H, La, Sr, Co, and O indicated in black, green, orange, blue, and pink, respectively.

Bader charge analysis was performed for the same set of DFT calculations to understand the electron distribution as a function of the host ($\delta = 0, 0.25, 0.5$) and $y$. The average Co Bader charge across all three magnetic initializations is shown in **Figure 7b**. For the unprotonated phases, the Co Bader charge decreases across the P-to-BM transition, consistent with Co reduction. Similarly, all three host phases show decreasing Co Bader charge with increasing $y$. The Co Bader charge is fairly consistent across different compositions with the same nominal Co oxidation state, indicating that electrons from $v_O$ and $H_i$ are approximately equally likely to localize on nearby Co. This suggests that the difference in $v_O$ and $H_i$ defect formation energies below the Co reduction limit stems from factors beyond the extent of Co reduction. For example, the localized strain energy penalty of $v_O$ formation is likely higher than that of $H_i$. Beyond the reduction limit, the Co Bader charges decrease more modestly or stabilize, which is consistent with the formation of charge-neutral $H_2$ dimers at high hydrogen insertion concentrations.

The structures and Bader charges corresponding to the DOS in **Figures 7c** and **7d** are shown in **Figure S7** and **Table S2**, respectively. Examining the protonated structures, significant distortion of the $CoO_6$ octahedra is observed for P $H_{1.5}La_{0.5}Sr_{0.5}CoO_3$, with many of the Co significantly offset from the centers of the octahedra. For BM $H_{0.5}La_{0.5}Sr_{0.5}CoO_{2.5}$, the structure is more well-preserved, with minimal distortions to the $CoO_6$ and $CoO_4$ polyhedra. For both structures, H forms hydroxyl bonds with O. Additionally, the UMA-identified ground-state $H_i$ configuration for BM finds all these hydroxyl bonds to occur at the apical O sites, in good agreement with prior computational findings that these are the most favorable bindings sites in BM SCO. [40] The Bader charges for all four structures confirm that H is positively charged and that La/Sr do not exhibit significant redox activity (**Table 2**).

These findings demonstrate that structural expansion and an increasing band gap are the consequences of $H_i$ formation across the P-to-BM transition, with the inserted hydrogen forming hydroxyl bonds and reducing nearby Co. For the compositions nominally corresponding to the reduction limit of Co, there are nearly zero O *2p* states and very few Co *3d* states at the lower energies of the conduction band, highlighting the need for a different redox mechanism, such as $H_2$ dimer formation, to accommodate further insertion of hydrogen.

**IV. Conclusion**

This study assesses the tendency and consequences of protonation in $La_{0.5}Sr_{0.5}CoO_{3-\delta}$ across the perovskite ($\delta = 0$) to brownmillerite ($\delta = 0.5$) transition. Dilute defect calculations demonstrate that oxygen vacancy formation energies increase drastically across the transition, while hydrogen interstitial formation energies are negative and only vary slightly for chemical potential conditions consistent with $p_{O_2}$ and $p_{H_2}$ of 1 bar at 300 K. This indicates facile insertion of hydrogen across the P-to-BM transition if hydrogen is available in the environment. Comparison of the $v_O$ and $H_i$ formation energies shows $H_i$ formation to be favored for a broad range of conditions at lower temperatures, such as those seen during EDLT device operation. At elevated temperatures, $v_O$ formation becomes more favorable across most conditions, implying that protonation will not interfere with the use of LSCO in high-temperature applications such as thermochemical water-splitting or solid oxide fuel cells. Non-dilute defect calculations yield insights into the limiting factors of protonation, revealing negative or near-zero interaction energies between $H_i$ up to the Co reduction limit ($Co^{2+}$) of the host phase ($\delta = 0, 0.25$, or $0.5$). Above this limit, $H_i$ formation energies increase rapidly with the concurrent observation of $H_2$ dimers instead of further hydroxyl (-OH) bond formation. Non-dilute $H_i$ concentrations also induce significant structural expansion and distortion to minimize strain energy. Although $H_i$ formation is favorable across the P-to-BM transition according to defect energetics, a convex hull analysis with respect to other compounds in the H-La-Sr-Co-O phase space reveals that all $H_yLa_{0.5}Sr_{0.5}CoO_{3-\delta}$ ($y = 0$ to 2, $\delta = 0, 0.25, 0.5$) phases are unstable towards decomposition. With this instability increasing with $y$, our findings suggest that hydrogen insertion will be facile during EDLT cycling, but the resulting protonated phases will possess a strong driving force towards decomposition, consistent with the acid-etching observed experimentally. Since the kinetics of hydrogen insertion are likely faster than those of phase decomposition, which requires complete bond breaking, it may be possible to access metastable protonated phases of LSCO, as has been shown for BM $HSrCoO_{2.5}$. Based on our calculations, these metastable structures are expected to exhibit considerable structural expansion, reduction of Co, and widening of the band gap. Importantly, these effects all mirror those of oxygen vacancy formation, making it difficult to indirectly confirm or reject protonation experimentally. Taken together, our results highlight protonation as an overlooked but important consideration when interpreting experimental results.

## Acknowledgements

This work was supported primarily by the National Science Foundation through the University of Minnesota MRSEC under Award Number DMR-2011401. This material is based upon work partially supported by the National Science Foundation Graduate Research Fellowship Program under Grant No. 2237827. Any opinions, findings, and conclusions or recommendations expressed in this material are those of the author(s) and do not necessarily reflect the views of the National Science Foundation. The authors acknowledge the Minnesota Supercomputing Institute (MSI) at the University of Minnesota for providing resources that contributed to the research results reported within this paper. The authors acknowledge Rohan Chakraborty, Vivian Ferry, Chris Leighton, Jierui Liang, and Kelsey Stoerzinger for helpful discussions related to this work.

28

# Supplementary Information

# Thermodynamics of proton insertion across the perovskite-brownmillerite transition in $La_{0.5}Sr_{0.5}CoO_{3-\delta}$

Armand J. Lannerd[a], Nathan J. Szymanski[a], and Christopher J. Bartel[a,*]

[a]University of Minnesota, Department of Chemical Engineering and Materials Science, Minneapolis, MN 55455

*correspondence to cbartel@umn.edu

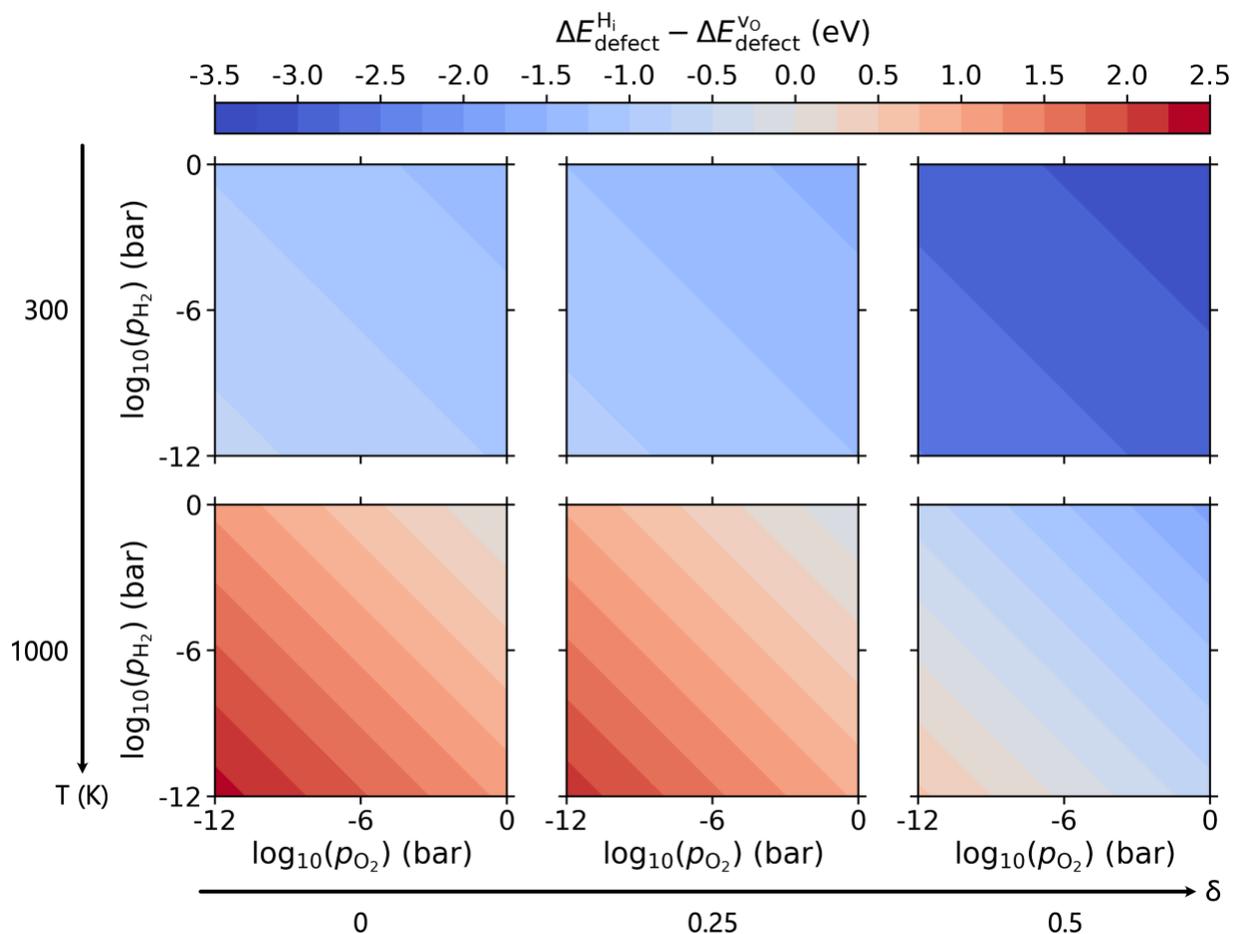

**Figure S1.** Defect competition between dilute oxygen vacancy ($v_O$) and dilute hydrogen interstitial ($H_i$) formation across the topotactic transition of $La_{0.5}Sr_{0.5}CoO_{3-\delta}$ from perovskite ($\delta = 0$) to brownmillerite ($\delta = 0.5$). Dilute formation energies ($\Delta E_{defect}$) were determined using DFT. Blue shading indicates conditions where dilute $H_i$ formation has a lower energy than dilute $v_O$ formation, whereas red indicates the opposite. All dilute defect formation energies were calculated with no applied electrochemical potential. The first row shows the defect competition at 300 K, while the second row denotes the competition at 1000 K. The lower left of each subplot reflects ultra-high vacuum (or ultra-low concentration) conditions for oxygen and hydrogen, while the upper right reflects higher pressure (concentration) conditions up to 1 bar. Analogous calculations performed at 300 K, but with an applied electrochemical potential are presented in **Figure 3**.

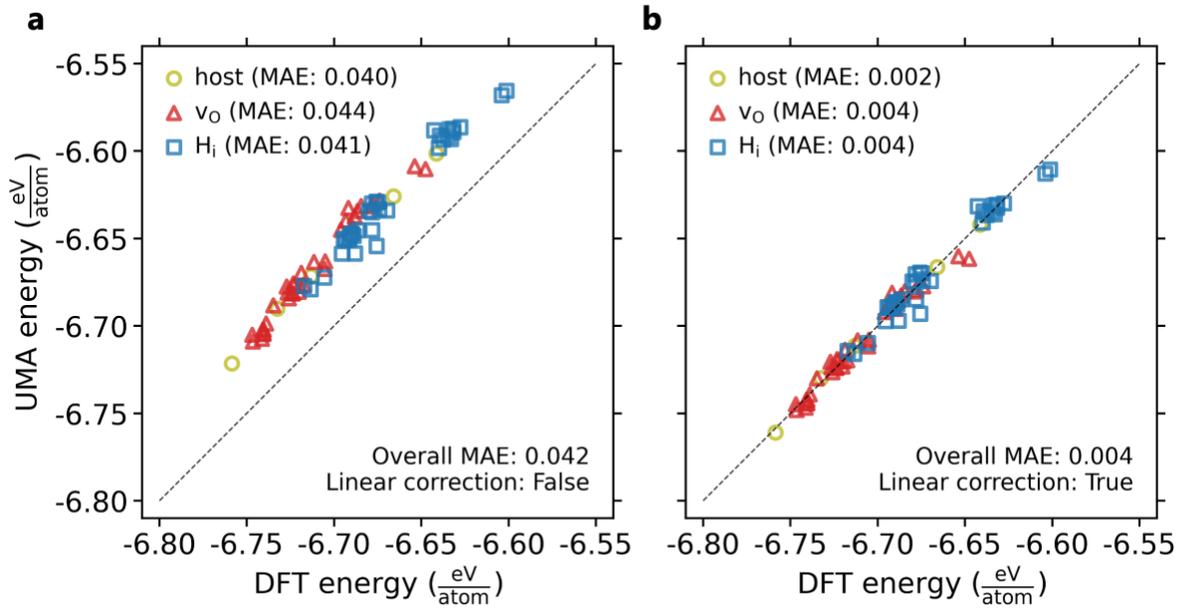

**Figure S2. a)** Parity plot comparing total energies from UMA [1] and DFT for host and dilute oxygen vacancy ($v_O$) and hydrogen interstitial ($H_i$) structures in $La_{0.5}Sr_{0.5}CoO_{3-\delta}$ for $\delta$ = 0, 0.125, 0.25, 0.375, and 0.5. **b)** Parity plot for the same data as in (a), except a linear correction has been applied to remove systematic error.

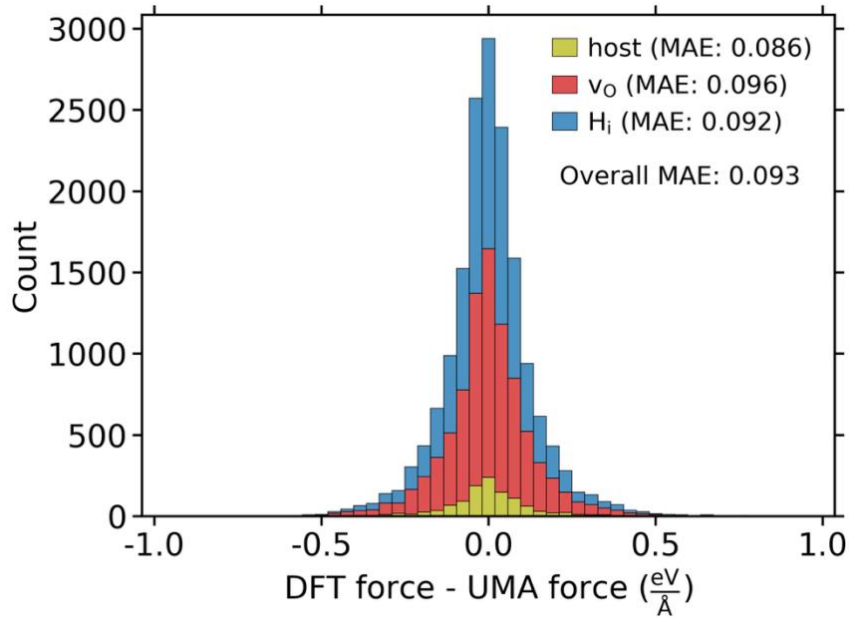

**Figure S3.** Error distribution and mean absolute errors (MAE) of UMA [1] vs DFT-predicted forces (for each atom and Cartesian component) for the converged host and dilute oxygen vacancy

($v_O$) and hydrogen interstitial ($H_i$) structures in $La_{0.5}Sr_{0.5}CoO_{3-\delta}$ for $\delta = 0, 0.125, 0.25, 0.375$, and $0.5$ (5 host, 35 $v_O$, and 35 $H_i$ structures).

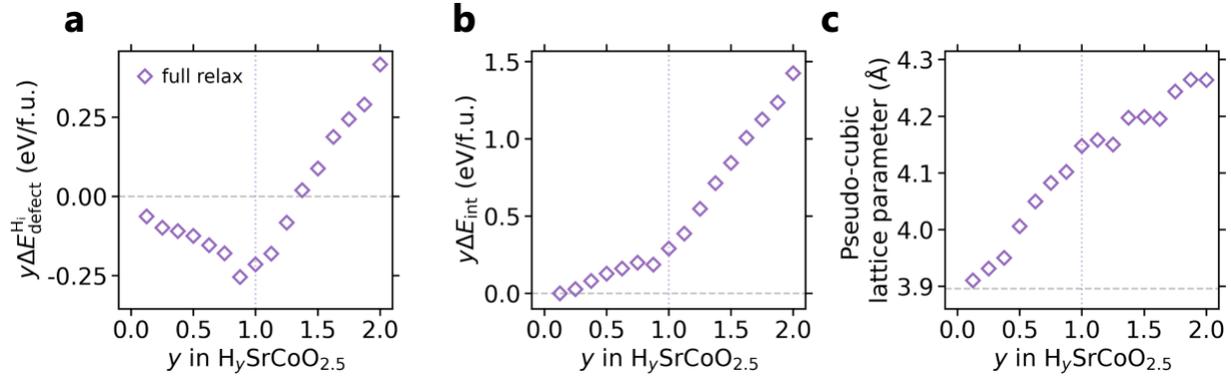

**Figure S4.** UMA [1] calculations of non-dilute hydrogen interstitials ($H_i$) in brownmillerite $SrCoO_{2.5}$ for comparison to DFT calculations from ref [2]. **a)** $H_i$ formation energies ($y\Delta E_{\text{defect}}^{H_i}$) relative to the pristine (no $H_i$) host for varying concentrations of $H_i$. Horizontal grey dashed lines correspond to defect formation energies of 0 eV/f.u. **b)** Interaction energies ($y\Delta E_{\text{int}}$) between $H_i$ for varying concentrations of $H_i$. Horizontal dashed lines correspond to defect interaction energies of 0 eV/f.u. **c)** Pseudo-cubic lattice parameter tracking structural expansion with increasing concentration of $H_i$. Horizontal grey dashed lines correspond to the pseudo-cubic lattice parameter of the pristine host. For all calculations, the chemical potential conditions are consistent with $p_{H_2}$ of 1 bar at 300 K. In these plots, $\Delta E_{\text{defect}}^{H_i}$ and $\Delta E_{\text{int}}$ are on a per defect basis. Multiplying them by $y$ (the molar composition of hydrogen per formula unit) converts them to a per formula unit basis to maintain a consistent scale. Across the subplots, the vertical dotted line corresponds to compositions where Co nominally has an oxidation state of 2+.

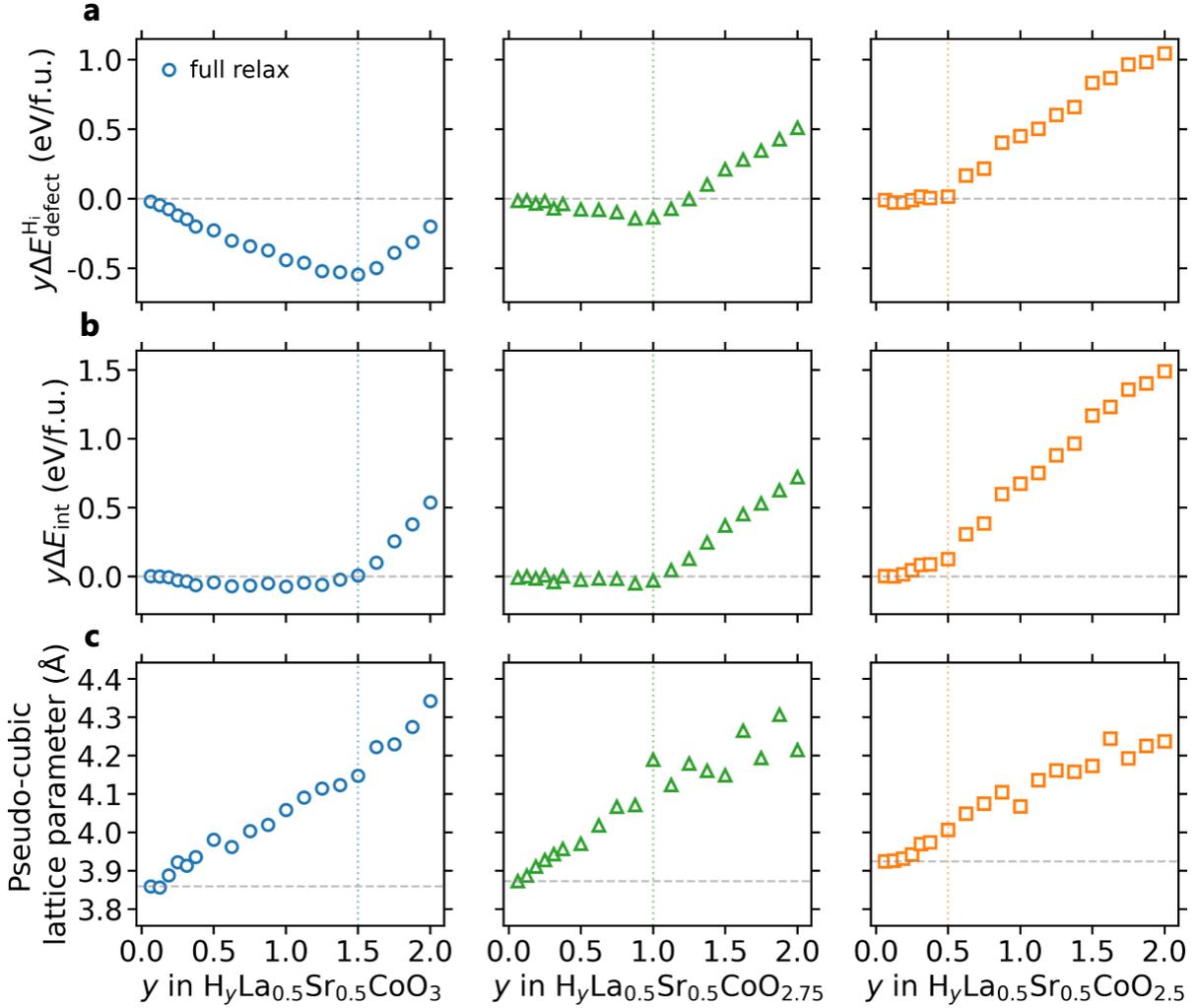

**Figure S5.** DFT calculations of non-dilute hydrogen interstitials ($H_i$) across the topotactic transition of $La_{0.5}Sr_{0.5}CoO_{3-\delta}$ from perovskite ($\delta = 0$) to brownmillerite ($\delta = 0.5$). **a)** $H_i$ formation energies ($y\Delta E_{defect}^{H_i}$) relative to the pristine (no $H_i$) host for varying concentrations of $H_i$. Horizontal grey dashed lines correspond to defect formation energies of 0 eV/f.u. **b)** Interaction energies ($y\Delta E_{int}$) between $H_i$ for varying concentrations of $H_i$. Horizontal dashed lines correspond to defect interaction energies of 0 eV/f.u. **c)** Pseudo-cubic lattice parameter tracking structural expansion with increasing concentration of $H_i$. Horizontal grey dashed lines correspond to the pseudo-cubic lattice parameter of the pristine host. For all calculations, the chemical potential conditions are consistent with $p_{H_2}$ of 1 bar at 300 K. In these plots, $\Delta E_{defect}^{H_i}$ and $\Delta E_{int}$ are on a per defect basis. Multiplying them by $y$ (the molar composition of hydrogen per formula unit) converts them to a per formula unit basis to maintain a consistent scale. Across the subplots, vertical dotted lines correspond to compositions where Co nominally has an oxidation state of 2+. Blue circles in the first column pertain to the perovskite ($\delta = 0$) host structure, green triangles in the second column the $\delta = 0.25$ host structure, and orange squares in the third column the brownmillerite ($\delta =$

0.5) host structure. In these calculations, both atom positions and lattice parameters were fully relaxed. The corresponding UMA [1] calculations are presented in **Figure 4**.

**Table S1.** Comparison of the lowest energy dilute oxygen vacancy ($v_O$) and dilute hydrogen interstitial ($H_i$) formation energies ($\Delta E_{\text{defect}}$) calculated via UMA [1] and DFT across the topotactic transition of La$_{0.5}$Sr$_{0.5}$CoO$_{3-\delta}$ from perovskite ($\delta = 0$) to brownmillerite ($\delta = 0.5$).

| Host composition | $\Delta E_{\text{defect}}^{v_O}$ (eV) | | $\Delta E_{\text{defect}}^{H_i}$ (eV) | |
|---|---|---|---|---|
| | UMA | DFT | UMA | DFT |
| La$_{0.5}$Sr$_{0.5}$CoO$_3$ | 1.304 | 1.081 | -0.413 | -0.302 |
| La$_{0.5}$Sr$_{0.5}$CoO$_{2.75}$ | 1.353 | 1.373 | -0.169 | -0.236 |
| La$_{0.5}$Sr$_{0.5}$CoO$_{2.5}$ | 3.015 | 3.038 | -0.130 | -0.168 |

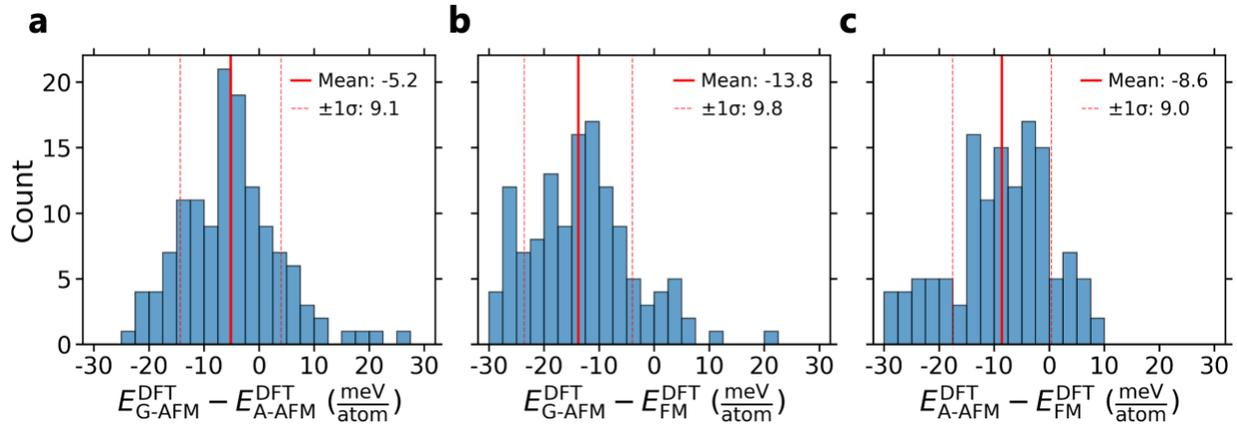

**Figure S6.** Distributions of difference in total energies ($E^{\text{DFT}}$) between **a)** G-AFM and A-AFM, **b)** G-AFM and FM, and **c)** A-AFM and FM magnetic moment initialization for DFT calculations across host, dilute oxygen vacancy, and non-dilute ($y = 0.0625$ to 2) hydrogen interstitial structures in H$_y$La$_{0.5}$Sr$_{0.5}$CoO$_{3-\delta}$ ($\delta = 0, 0.25, 0.5$). Vertical red bars denote the mean (solid) and one standard deviation (dashed) of each distribution. The ground-state (lowest energy) result followed from G-AFM initializations in 72.5% of cases, A-AFM in 22.9%, and FM in 4.6%.

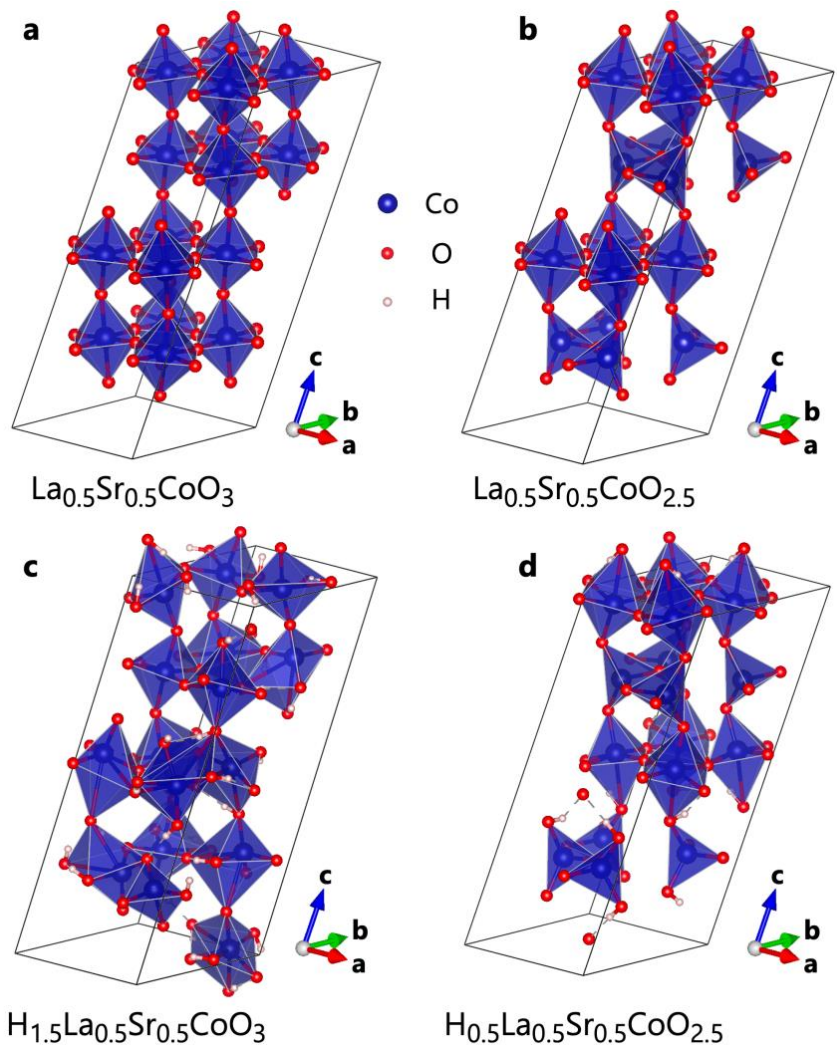

**Figure S7.** DFT-relaxed structures of **a)** $La_{0.5}Sr_{0.5}CoO_3$, **b)** $La_{0.5}Sr_{0.5}CoO_{2.5}$, **c)** $H_{1.5}La_{0.5}Sr_{0.5}CoO_3$, and **d)** $H_{0.5}La_{0.5}Sr_{0.5}CoO_{2.5}$ corresponding to the DOS from **Figure 7**. The *A*-site La and Sr are hidden for clarity, and Co, O, and H are represented by blue, red, and white spheres, respectively.

**Table S2.** Element-wise Bader charges corresponding to the DOS in **Figure 7** and the structures in **Figure S7**. The average H, Co, and O Bader charges are bolded for clarity.

| Host composition | H | | La | | Sr | | Co | | O | |
|---|---|---|---|---|---|---|---|---|---|---|
| | Avg | Std | Avg | Std | Avg | Std | Avg | Std | Avg | Std |
| $La_{0.5}Sr_{0.5}CoO_3$ | - | - | 2.097 | 0.005 | 1.597 | 0.003 | **1.451** | 0.048 | **-1.099** | 0.018 |
| $La_{0.5}Sr_{0.5}CoO_{2.5}$ | - | - | 2.059 | 0.004 | 1.578 | 0.004 | **1.260** | 0.107 | **-1.232** | 0.031 |
| $H_{1.5}La_{0.5}Sr_{0.5}CoO_3$ | **0.576** | 0.018 | 2.078 | 0.038 | 1.595 | 0.025 | **1.161** | 0.064 | **-1.287** | 0.028 |
| $H_{0.5}La_{0.5}Sr_{0.5}CoO_{2.5}$ | **0.552** | 0.014 | 2.041 | 0.012 | 1.574 | 0.010 | **1.148** | 0.016 | **-1.293** | 0.018 |